\documentclass{ethpaper}
\usepackage{graphicx}
\usepackage{subfigure}
\usepackage{amssymb}
\begin{document}
\begin{titlepage}
\ethnote{}
\title{Comparison between high-energy proton and charged pion induced damage \\
in PbWO${_4}$ calorimeter crystals}
\begin{Authlist}
P.~Lecomte, D.~Luckey, F.~Nessi-Tedaldi, F.~Pauss
\Instfoot{eth}{Institute for Particle Physics, ETH Zurich, 8093 Zurich, Switzerland}
D.~Renker
\Instfoot{psi}{Paul Scherrer Institute, Villigen PSI, 5232 Villigen, Switzerland}
\end{Authlist}
\maketitle

\begin{abstract}

  A PbWO$_4$ crystal produced for the electromagnetic calorimeter of
  the CMS experiment at the LHC was cut into three equal-length
  sections. The central one was irradiated with 290 MeV/c positive
  pions up to a fluence of $(5.67 \pm 0.46)\times 10^{13}\;
  \mathrm{cm^{-2}}$, while the other two were exposed to a 24\,GeV/c
  proton fluence of $(1.17 \pm 0.11) \times
  10^{13}\;\mathrm{cm^{-2}}$.  The damage recovery in these crystals,
  stored in the dark at room temperature, has been followed over two
  years. The comparison of the radiation-induced changes in light
  transmission for these crystals shows that damage is proportional to
  the star densities produced by the irradiation.

\end{abstract}

\vspace{7cm}
\conference{submitted to Elsevier for publication in Nucl. Instr. and Meth. in Phys. Research A}

\end{titlepage}
\section{Introduction}
\label{s-int}
A recent study we performed on Lead Tungstate crystals has
demonstrated that hadrons cause a specific, cumulative damage which
only affects light transmission, while the scintillation mechanism
remains unaffected~\cite{r-LTNIM,r-LYNIM}. The results were obtained
exposing the crystals to various fluences of 20 GeV/c or 24 GeV/c
proton beams up to the full integrated fluence expected at the
LHC. Complementary $\gamma$ irradiations with a $^{60}$Co source
allowed to disentangle the damage due to the associated
ionising dose.

However, crystals used in high-energy physics detectors will typically
be exposed to hadrons -- mostly charged pions -- with different
energies. In the CMS experiment at the LHC for example~\cite{r-TDR},
the large hadron fluxes are due to particles whose energies rarely
exceed 1 GeV.  Thus, it had to be established how results obtained
with high-energy protons can be scaled to lower energies and different
particle types.

\section{The crystals}
\label{s-cry}
For this study, a $\mathrm{PbWO}_4$ crystal was used, labelled {\em w}
in Refs.~\cite{r-LTNIM} and \cite{r-LYNIM}, produced by the
Bogoroditzk Techno-Chemical Plant (BTCP) in Russia for the
electromagnetic calorimeter (ECAL) of the CMS experiment~\cite{r-TDR}.
This crystal had been already tested under irradiation with $^{60}$Co
photons up to 9.59 kGy, showing a modest induced absorption at the
peak of scintillation-emission wavelength,
$\mu_{IND}\mathrm{(420\;nm)}\simeq 0.2\; $m$^{-1}$.  The damage from
$\gamma$ irradiation was annealed by heating the crystal to $250^o$ C
for 4 h and full recovery was checked through light transmission
measurements. Then, the crystal was cut into three equal-length
sections, with nearly parallelepipedic dimensions of $2.4 \times 2.4\;
\mathrm{cm}^2$ and lengths of 7.5 cm, which we labelled {\em w1}, {\em
  w2} and {\em w3}.  The first, {\em w1}, and last, {\em w3}, sections
were irradiated with 24\,GeV/c protons at the IRRAD1
facility\,\cite{r-IR1} in the T7 beam line of the CERN PS accelerator,
while the middle section, {\em w2}, was irradiated with 290 MeV/c
pions in the $\pi E1$ beam line at the Paul Scherrer Institute (PSI)
in Villigen, Switzerland.
\section{The proton irradiation}
\label{s-PRI}
Samples {\em w1} and {\em w3} were irradiated at the same time,
with {\em w3} placed right behind {\em w1}, so that the proton beam
was entering through the small face of {\em w1} and the hadronic
cascade could develop through both crystals.  The same irradiation
procedure was followed as in \cite{r-LTNIM}, where all
details can be found.  The fluence was determined through
the activation of an aluminium foil covering the crystal front face.
The proton beam spot was broadened to cover the whole crystal front
face with a flux $\phi_p= 2.8 \times 10^{12}$\,cm$^{-2}$h$^{-1}$. The
proton fluence reached was $\Phi_p=(1.17 \pm 0.11) \times
10^{13}\;\mathrm{cm^{-2}}$.
\section{The pion irradiation}
\label{s-PII}
The pion irradiation was performed in the high-flux secondary pion
beam line $\pi E1$ at the Paul Scherrer Institute 590 MeV Ring
Cyclotron~\cite{r-SUG}. 

Pions are produced there by primary protons hitting a graphite target. They
are then extracted from the target at an angle of 8$^{o}$ with respect
to the incident protons and then transferred by a beam line containing
a magnet-spectrometer in order to select them according to charge and
momentum. The beam line was set to deliver positively charged pions to
the irradiation zone at a nominal momentum of 300 MeV/c.  The protons
and positrons contamination was suppressed by inserting 6 mm and 15 mm
Carbon foils respectively before and after the last bending
magnet. The resulting beam momentum on the crystal was $(290.2 \pm
0.3)$ MeV/c, where the error is dominated by the uncertainty in the
energy loss in carbon.  The neutrons produced by proton interactions
in the graphite foils yield a beam contamination below
1\%\cite{r-FUR}, and the positron contamination is of the order of
0.5\%\cite{r-REN}.

To uniformly irradiate the crystal, it was longitudinally positioned
at the waist of the beam in the irradiation zone, where the
contribution of the divergence to the beam spot is minimal.
Transversally, the beam spot was optimised to have a nearly Gaussian
shape in both transverse directions, with a FWHM of $\sim 42$ mm both,
in the vertical and in the horizontal directions.  The beam profile
was monitored with an X-Y wire chamber placed 11 cm upstream of the
crystal, and whose wire signals were displayed on an oscilloscope.  A
ionization chamber, also placed in the beam, was used to monitor the
beam fluence: its digitised induced current $N_{ICS}$, which is
proportional to the total beam intensity, was integrated throughout
the irradiation.

The crystal was placed on a 20 cm thick Styrofoam support to minimise
the amount of surrounding material.  Before the start of irradiation,
the beam spot profiles were checked in 3 planes by means of
self-developing Gafchromic MD55 dosimetry foils~\cite{r-FOI}. The
foils were placed longitudinally at the coordinates corresponding to
the entrance face, middle and exit face of the crystal, supported by a
Styrofoam holder.  A 10 min long exposure of the foils to the beam
provided a sufficient contrast to visualise uniform, equally sized
beam spots in the three positions.  The pion fluence measurements were
performed using the activation of aluminium foils, as described in
section \ref{s-PIF}. Further details about the beam line setup and
beam control can be found in Ref. \cite{r-SUG}, \cite{r-REN} and
\cite{r-SUH}. The crystal was irradiated for 137.4 h, for a total
fluence $\Phi_{\pi}=(5.67 \pm 0.46)\times 10^{13}\;
\mathrm{cm^{-2}}$. The average flux on the crystal was
$\phi_{\pi}=4.13\times 10^{11}\; \mathrm{cm^{-2}\; h^{-1}}$.
\section{Pion fluence determination}
\label{s-PIF}
For the pion fluence determination we used the activation of aluminium
foils by the beam~\cite{r-FUR}, by determining, with a Germanium
spectrometer~\cite{r-MAL}, the amount of $^{22}$Na or $^{24}$Na
isotopes present at the end of irradiation.  For this purpose, we
placed a 1.588 g aluminium foil, $2.4 \times 2.4\; \mathrm{cm}^2$ in
cross-section, 1 cm upstream of the crystal.  For the crystal
irradiation, which lasted much longer than the $^{24}$Na lifetime
$\tau_{24}= 21.6$ h, $^{22}$Na was more suitable for a fluence
determination. However, precise values of the production cross section
for the considered pion energy range can be found in literature only
for $^{24}$Na~\cite{r-24NA}. From the existing data we determined an
interpolated value at our beam energy, which amounts to
\begin{equation}
\sigma\left(A\ell(\pi^+,X)^{24}Na\right)=(20.0 \pm 0.7)\; \mathrm{mb}.
\label{e-sigma24}
\end{equation}
In order to use the $^{22}$Na activation for a precise fluence
determination for the crystal irradiation, we measured the
$^{24}$Na/$^{22}$Na cross section ratio through a 12 h long activation
of a 6.3173 g aluminium foil, $2.4 \times 2.4\; \mathrm{cm}^2$ in
cross-section, exposed to the pion beam without the presence of a
crystal. Furthermore, during all the pion irradiations, the
instantaneous primary beam intensity was recorded every 5
s~\cite{r-MEZ} .

The pion fluence on the foil can be calculated from the number
$K_{24}$ of created $^{24}$Na nuclei, using the known 
$\sigma\left(A\ell(\pi^+,X)^{24}Na\right)$ cross section (Eq.\ref{e-sigma24}),
which we label $\sigma_{24}$:
\begin{equation}
K_{24} = \kappa \cdot \sigma_{24} \sum_{i=1}^{n} I_i\cdot \Delta t_i
\end{equation}
with $\kappa$ a proportionality constant  
and $I_i$ the beam intensity for a time interval $\Delta t_i\;
(1 \leq i \leq n)$, provided that $\Delta t_i \ll \tau_{24}$.

However, due to the isotope decay, the measured activity
$\mathcal{A}_{24}$ for the $A\ell$ foil at the time the irradiation
ended, $t_{END}$, is the one of the leftover isotopes, and it is given
by:
\begin{equation}
\mathcal{A}_{24} = \frac{\kappa \cdot \sigma_{24}}{\tau_{24}} \sum_{i=1}^{n} I_i\cdot e^{-(t_{END}-t_i)/\tau_{24}}\Delta t_i .
\end{equation}
Thus, from a measurement of $\mathcal{A}_{24}$ and a precise knowledge of the 
instantaneous beam intensities  $I_i$ throughout the irradiation,
it was possible to calculate the true amount $K_{24}$ of created isotopes:
\begin{equation}
K_{24} = \mathcal{F}_{24}\cdot \mathcal{A}_{24} \cdot \tau_{24}
\end{equation}
where 
\begin{equation}
\mathcal{F}_{24} = \frac{\sum_{i=1}^{n} I_i\cdot \Delta t_i}{\sum_{i=1}^{n} I_i\cdot e^{-(t_{END}-t_i)/\tau_{24}}\Delta t_i}.
\end{equation}
The uncertainty $\Delta \mathcal{F}_{24}$ on $\mathcal{F}_{24}$ was calculated as
\begin{equation}
(\Delta \mathcal{F}_{24})^2 = \sum_{i=1}^{n} \left( \frac{\partial \mathcal{F}_{24}}{\partial I_i}\right) ^2(\Delta I_i)^2
\end{equation}
where each $\Delta I_i$ was taken as the half excursion between two
subsequent intensity values.  The average primary beam intensity was
calculated as
\begin{equation}
\overline{I} = \frac{\sum_{i=1}^{n} I_i\cdot \Delta t_i}{\sum_{i=1}^{n}\Delta t_i}.
\label{e-INT}
\end{equation}
and its uncertainty similarly to the one for $\mathcal{F}_{24}$. For
the irradiation of the $A\ell$ foil, we obtained $\overline{I} = (
1744.35 \pm 1.76) \;\mu$A and $\mathcal{F}_{24} = 1.249 \pm 0.001$,
while during the crystal irradiation we had $\overline{I} = ( 1653.58
\pm 0.21)\; \mu$A.  The small uncertainties in average beam
intensities show how stable the beam conditions were throughout the
irradiation.  The determination of the $^{22}$Na activity proceeded
analogously, but no
corrections for decays during the irradiation needed to be applied,
since its duration was much shorter than the $^{22}$Na life time,
$\tau_{22} = 2.6$ y.

After corrections for beam intensity fluctuations and for the isotope
decays since the time of production, the spectrometric analysis of the
foil irradiated without a crystal yields a cross section ratio
\begin{equation}
\frac{\sigma\left(A\ell(\pi^+,X)^{24}Na\right)}{\sigma\left(A\ell(\pi^+,X)^{22}Na\right)}= \frac{K_{24}}{K_{22}}=0.785 \pm 0.048
\label{e-RNa}
\end{equation}
and thus, using Eq.~\ref{e-sigma24},
\begin{equation}
\sigma\left(A\ell(\pi^+,X)^{22}Na\right)=(25.5 \pm 1.8)\; \mathrm{mb}
\end{equation}
This
cross section, together with the spectrometric $^{22}$Na activity
analysis applied to the foil placed in front of the crystal yields,
assuming that all $^{22}$Na is due to activation by pions, a total
pion fluence $\Phi_{\pi}=(5.67 \pm 0.46)\times 10^{13}\;
\mathrm{cm^{-2}}$ for the crystal irradiation.
 
While beam contaminations by other particles are
negligible~\cite{r-REN,r-GLA}, $^{22}$Na production by neutrons
originating from the hadron cascade in the crystal had to be
considered. To quantify this systematic effect, we determined the
ratio of the incoming beam fluences measured by the ionization chamber
during the foil activation and during the crystal irradiation,
  \begin{equation}
  R_{ICS}=\frac{N_{ICS}^{A\ell + Crystal}}{N_{ICS}^{A\ell}}
  \end{equation}
  and compared it to the ratio of fluences, $R_{Na}$, determined from
  the $^{22}$Na activation of the foils assuming it was all due to
  beam particles. We obtain
  \begin{equation}
   R_{ICS}=13.16\pm 0.40
  \end{equation}
to be compared with
\begin{equation}
   R_{Na}=13.44 \pm 0.76.
  \end{equation}
  The two ratios are consistent within the precision of the
  measurements, which means that the amount of $^{22}$Na isotopes created by
  neutrons coming from the crystal can be neglected.
As an additional cross-check, the fluence ratios
  above can be compared to the fluence ratio of the primary beam for
  the two irradiations, which we obtain from Ref.~\cite{r-MEZ} data using Eq.~\ref{e-INT}:
 \begin{equation}
  R_{Beam}=13.02 \pm 0.01
  \end{equation}
  The consistency observed demonstrates that the beam has been very
  stable, and that the instantaneous primary beam intensities can be
  used to take into account the $^{24}$Na activity decay in $A\ell$
  foils during irradiation. 
  \section{Measurements and Results}
Hadrons in the range of energies and fluences considered, only change
 a crystal's light transmission, while the scintillation
 mechanisms remain unaffected, as we have shown in
 Ref.~\cite{r-LYNIM}. Thus, this study of damage focuses on the
 hadron-induced light transmission changes, which have been measured
 with a Perkin Elmer Lambda 900 spectrophotometer over the range of
 wavelengths between 300 nm and 800 nm, in steps of 1 nm.
\begin{figure}[bh]
\begin{center}\footnotesize
\mbox{\includegraphics[width=15cm]{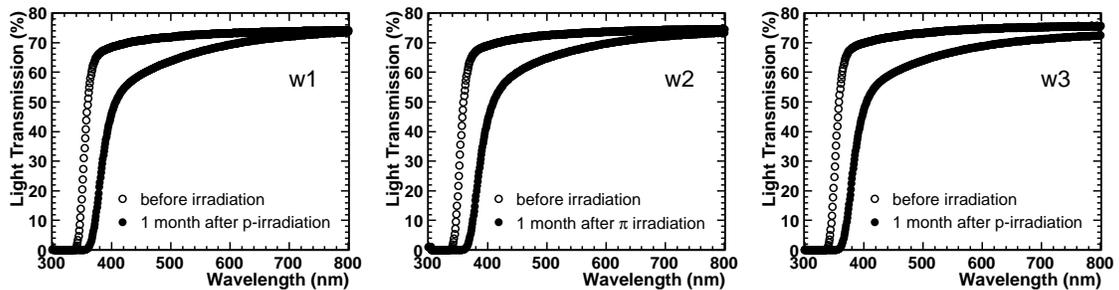}}
\end{center}
\caption{Longitudinal Light Transmission  curves for the three crystals
showing their degree of hadron-induced damage one month after the irradiation,
compared to the values before irradiation.
\label{f-LT}}
\end{figure}
\begin{figure}[th]
\begin{center}\footnotesize
\mbox{\includegraphics[width=15cm]{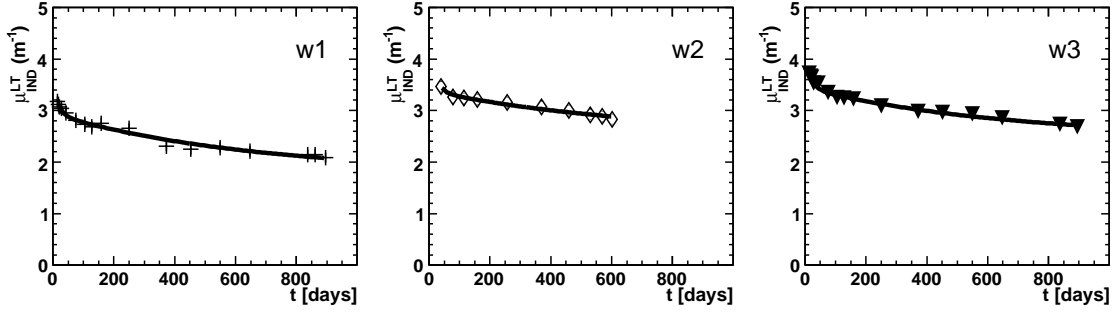}}
\end{center}
\caption{Recovery data for crystals
{\em w1} and {\em w3} after proton
irradiation, and for crystal {\em w2} after pion irradiation.
\label{f-REC}}
\end{figure}
\begin{figure}[bh]
\begin{center}\footnotesize
\mbox{\includegraphics[width=12cm]{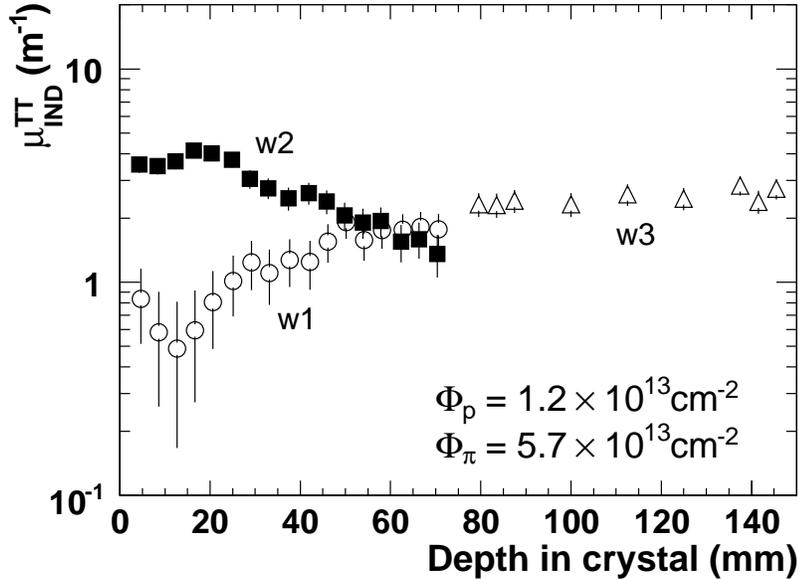}}
\end{center}
\caption{Induced absorption coefficient, $\mu_{IND}^{TT}(420\; \mathrm{nm})$,
measured transversely, as a function of position along
the crystals length, 150 days after irradiation. The {\em w1} and {\em w3}
data are placed according to the crystals' position during irradiation.
\label{f-MUT}}
\end{figure}
\begin{figure}[th]
\begin{center}\footnotesize
\mbox{\includegraphics[width=12cm]{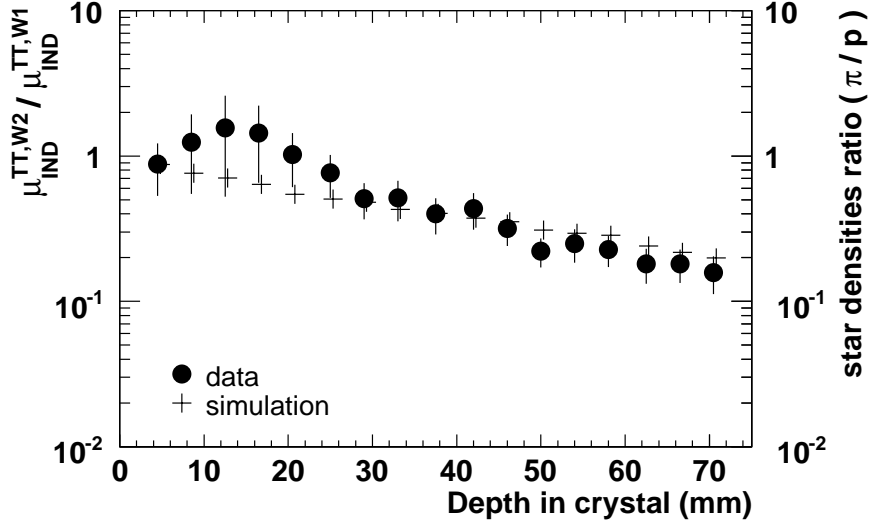}}
\end{center}
\caption{Ratio of induced transverse absorption coefficients for
{\em w2} and {\em w1} (black dots)
normalised to the same fluence, compared to the ratio of
star densities produced by pions and protons (crosses).
\label{f-MUR}}
\end{figure}
 Details to
 the measuring technique and precision can be found in
 Ref.~\cite{r-LTNIM}.The Longitudinal Light Transmission (LT) values measured one month
after irradiation on all three crystals through their 7.5 cm length are
shown in Fig.~\ref{f-LT}. 
It is evident that qualitatively, the LT
changes for crystal {\em w2} after pion irradiation are similar to the
ones for crystals {\em w1} and {\em w3} after the proton
irradiation. In particular, the shift in band-edge observed already in proton
irradiations is present in the $\pi$-irradiated {\em w2} crystal as
well, while it is absent in $\gamma$ irradiated ones~\cite{r-LTNIM}. The
comparable magnitude of damage was achieved on purpose by an
appropriate choice of irradiation fluences.

Light transmission was measured over time, to provide recovery data.
The damage is quantified through the induced absorption coefficient at
the peak of the scintillation emission, defined according
to~\cite{r-LTNIM} as:
\begin{equation}
\mu_{IND}^{LT}(420\; \mathrm{nm}) = \frac{1}{\ell}\times \ln \frac{LT_0}{LT}
\label{muDEF}
\end{equation}
where $LT_0\; (LT)$ is the Longitudinal Transmission value measured
before (after) irradiation through the length $\ell$ of the crystal.

The evolution of damage over time is shown in Fig.~\ref{f-REC} for all
three crystals. The data, taken over more than 2 years, are well
fitted, as in the proton irradiation studies of Ref.~\cite{r-LTNIM},
by a sum of a constant and two exponentials with time constants
$\tau_i\; (i=1,2)$:
\begin{equation}
\mu_{IND}^{LT,\; j}(420\; \mathrm{nm},t_{\rm{rec}}) = \sum_{i=1}^{2} A_i^je^{-t_{\rm{rec}}/\tau_i} + A_3^j,
\label{e-Afit}
\end{equation}
where $t_{\rm{rec}}$ is the time elapsed since the irradiation, while
$A_i^j,\; (i=1,2)$ and $A_3^j$ are the amplitude fit parameters for
crystal $j\; (j=1,2,3)$.  Figure \ref{f-REC} shows the results of a
fit where the recovery time constants have been kept fixed to the
values obtained in Ref.~\cite{r-LTNIM}, $\tau_1 = 17.2$ days and
$\tau_2 = 650$ days. A similar $\chi^2$ quality can be obtained
performing a fit of the form
\begin{equation}
\mu_{IND}^{LT,\; j}(420\; \mathrm{nm},t_{\rm{rec}}) = B_0^j\left(e^{-t_{\rm{rec}}/\tau_2}
                                        + B_2\right) + B_1^j e^{-t_{\rm{rec}}/\tau_1}.
\label{e-Bfit}
\end{equation}
Such a fit, where $B_2$ is forced to be the same for all three
crystals, corresponds, for $t_{\rm{rec}} >> \tau_1$, to a constant damage
amplitudes ratio between crystals, and makes further damage
comparisons independent from the time where the measurement is
performed.  The best fit is obtained from Eq.~\ref{e-Bfit}, and it
yields $\tau_1 = 52$ days, $\tau_2 = 369$ days and $B_2=3.75$. Since
$B_2$ corresponds to the ratio $\frac{A_3}{A_2}$ (see Eq.~\ref{e-Afit}),
the results indicate that 78\% of the long-term damage does not
recover, as it is also evident from Fig.~\ref{f-REC}.

To compare proton- and pion-damage, we performed measurements of the
Transverse Light Transmission (TT) profiles along the length of the
crystals, 150 days after irradiation, shining the spectrophotometer
light beam across their $\sim 2.4$ cm transverse dimension.
The date was chosen such that the $\tau_1$ component of the damage had
practically disappeared. The damage is quantified through the transverse
induced absorption coefficient at the peak of the scintillation
emission, $\mu_{IND}^{TT}(420\; \mathrm{nm})$, defined analogously to
$\mu_{IND}^{LT}(420\; \mathrm{nm})$ in Eq.~\ref{muDEF}, with $\ell$
the transverse crystal dimension for each given longitudinal position.
The measurements are shown in Fig.~\ref{f-MUT}. The damage as a
function of position
has the same shape as the star\footnote{A star is
  defined~\cite{r-LTNIM} as an inelastic hadronic interaction caused
  by a projectile above a given threshold energy.}  densities as a function
of depth obtained from FLUKA simulations (Fig.~3 in
Ref.~\cite{r-LTNIM} and Ref.~\cite{r-HUH}).  This constitutes yet
another confirmation of our understanding of the hadron damage
mechanisms in Lead Tungstate: it turns out to be proportional to the
star densities in the crystals, in that it is due to the very high
local ionization from fragments created in nuclear collisions.  The
observed decrease of damage with depth for crystal {\em w2} is due to
the absorption in the crystal and agrees with the measured $\pi$
absorption cross-section in Lead Tungstate that can be extracted from
Ref.~\cite{r-ASH}.

The present study was advocated in Ref.~\cite{r-LTNIM}, to
experimentally determine the factor needed to quantitatively scale the
damage measured for high-energy protons to the particle spectrum
expected at the LHC, which is mostly composed by pions with energies
$\leq 1$ GeV.  For this purpose, in Fig.~\ref{f-MUR} we have
normalised to the same particle fluence the transverse damage from
pions measured for crystal {\em w2} and the one from protons for
crystal {\em w1}, and we have plotted their ratio as a function of
depth.  It should be noticed there, that the large error bars for the
first 2 cm of depth are dominated by the uncertainties on the
measurement of the very small damage values from protons in {\em w1}.
In the same plot, we show the ratio of star densities obtained from
FLUKA simulations for the two cases~\cite{r-LTNIM,r-HUH}.  The
measured ratios and the star densities ratios are in agreement within
the experimental uncertainties. This demonstrates that, at least for
the considered particle types and energy range, the measured damage
can simply be rescaled to given experimental conditions through the
ratio of simulated star densities. Furthermore, in Ref.~\cite{r-LTNIM}
it was argued that, in addition to star densities, the total track
length of stars might play a role. The results shown in
Fig.~\ref{f-MUR} rule out this hypothesis.
\section{Conclusions}
\label{s-CON}
We have performed a 24 GeV/c proton irradiation of two PbWO$_4$
crystal samples and a 290 MeV/c $\pi^+$ irradiation of a third one, and we
have studied the damage caused to the crystal light transmission. The
longitudinal profile of the damage is proportional to the star
densities obtained from simulations, and the profile of induced
absorption coefficient ratios is well reproduced by the profile of
star densities ratios.  We conclude that, in a fluence regime where
the damage due to the associated ionising dose can be neglected, the
damage to be expected from pions at energies around 1 GeV/c can be
rescaled from the damage measured for 24 GeV/c protons by means of
star densities ratios obtained from simulations.
\section*{Acknowledgements}
We are indebted to R.~Steerenberg, who provided us with the required
CERN PS beam conditions for the proton irradiations. We are grateful
to the PSI accelerator staff for the very stable beam, and in
particular to A.-Ch.~Mezger for providing us also with detailed beam
intensity data.  We are deeply grateful to M.~Glaser and F.~Ravotti,
who helped us in operating the proton irradiation and dosimetry
facilities at CERN. We also gratefully acknowledge the help of
F.~Jaquenod and F.~Malacrida who performed the dosimetric measurements
at PSI after the pion irradiation.  M.~Huhtinen's contribution in
the early phases of preparatory work is warmly acknowledged.

\end{document}